# Impact of MQTT Based Sensor Network Architecture on Delivery Delay Time

Oleksandr Kovalchuk, Yuri Gordienko, Sergii Stirenko

National Technical University of Ukraine "Igor Sikorsky Kyiv Polytechnic Institute", Kyiv, Ukraine
me.olexandr.kovalchuk@gmail.com

**Abstract.** The purpose of this study is to present two new architectures of the complexes of embedded systems in the context of the Internet of Things. The proposed new schemes use the MQTT protocol, but provide both an autonomous operation and an online mode. This is achieved by adding one additional layer, which consists of its own MQTT-broker and critical messages handler, so that in case of the Internet access disappearance, the system could continue to respond to changes. The research of message delivery delays has been carried out for the proposed models. The comparative analysis was conducted to identify the advantages and disadvantages of both solutions.

## 1    Introduction

The rapid development of innovative technologies in the Internet of Things opens up new directions of modern research in the field of computer technology. A wide range of applications of these technologies and their effectiveness become powerful driving forces for the implementation of relevant innovations in almost all human activities.

The Internet of Things (IoT) is a network concept consisting of interconnected physical devices that have built-in sensors. An obligatory component is also the software that allows the transmission and exchange of data between the physical world and computer systems using standard communication protocols. In addition to sensors, the network may have actuators built into physical objects and interconnected via wired or wireless networks. These interconnected devices have the ability to read and operate on the data, identify themselves and be programmed. They also exclude the need for human involvement by the use of intelligent interfaces [1].

Internet of Things is also defined as a "smart environment", based on the constant proliferation of intelligent networks, wireless sensors and mass data centers [2]

The Internet of Things is also known as the Cloud of Things (CoT). In general, IoT and CoT differ from each other, but their features often complement each other. That is why many researchers advocate their integration in order to benefit from specific scenarios and in the wider context of Cloud-Fog-Dew computing paradigm [3-5].

An important direction of IoT development is the search for a convenient way of interacting these systems, which provides reliability, convenience and integration with many types of devices.

## 2 Background

MQTT, which stands for Message Queue Telemetry Transport, has proved itself as reliable protocol that has a small footprint on message data size. Despite not being new protocol, it is actively developed: latest version of protocol standard has been released in October, 2018 [6].

Currently the most widely used version of MQTT is 3.1.1. In comparison with the latest version of the protocol, it lacks some performance features like topic aliases, improved flow control and shared subscriptions as well as convenience features like user properties and enhanced authentication [6]. Several systems, which use MQTT of version 3.1.1 for communication, are investigated now. The version of the MQTT protocol, which was developed specially for the usage in sensor networks, is based on the MQTT 3.1.1 and called MQTT-SN. However, this one is not covered in the given paper.

MQTT is a publish/subscribe protocol, which means in systems based on this protocol there are producers, which publish data to the topics and consumers, which subscribe to the topics with the data they are interested in. There is also the broker entity in the system: MQTT server that receives messages from publishers and deliver them to corresponding subscribers based on the topics and quality of service policy. Publish/subscribe approach allows to decouple implementations of producer and consumer and add or replace components of the system as long as data format and topics names are preserved. This gives the flexibility to the system, as its components do not depend on each other's implementation, but only depend on data they provide [6].

The simplified scheme of interaction in the MQTT-based systems can be described as follows. Producer or publisher connects to MQTT broker. Consumers, which are also called subscribers connect to the same MQTT broker and subscribe to the topics they are interested in. When the producer has information to share, it publishes a message to the broker, which then shares the information to all subscribers of this topic.

Message Queue Telemetry Transport also provides several quality of service (QoS) policies, which provide different guarantees at the cost of delivery slowdown and network traffic increase. Slowdowns are minimal when using QoS 0 (at max once), however this policy does not guarantee that message will be delivered after it was sent. QoS 1 (at least once) guarantees that message will be delivered to the destination at least once, however duplicate messages may occur. The most strict policy is QoS 2, which guarantees exactly one message delivery, which come at cost of at least three additional messages being sent over network, and thus larger delays. The latter is used in the system, where delivery is critical, and message duplicates are not allowed [7].

In this investigation, QoS 1 is used as the policy, which provides optimal balance between delivery reliability and footprint it has for most of sensor networks.

## 3 Experimental and Computational Details

One of the ways to maintain the autonomy of the system, is to transfer part of business logic and communication to sensor network itself instead of using only public broker. In this case, the part of the functionality should still be present in the external network, so that it can be accessed from the outside of the sensor network (for example, for users to receive statistics and alerts on their smartphones). Having external brokers also allow more powerful computational resources to access sensor networks data for processing.

### 3.1. Proposed alternative architectures

According to the above, we propose two alternative architectures to consider. Both proposed architectures have an additional layer of MQTT broker and critical message handler inside the sensor networks. Proposed architectures allow keeping system operating even when Internet access is lost. First of the two proposed architectures has a lightweight copy of the whole system in every sensor network (Fig. 1). This allows keeping sensor nodes logic clean and handling network loss separately.

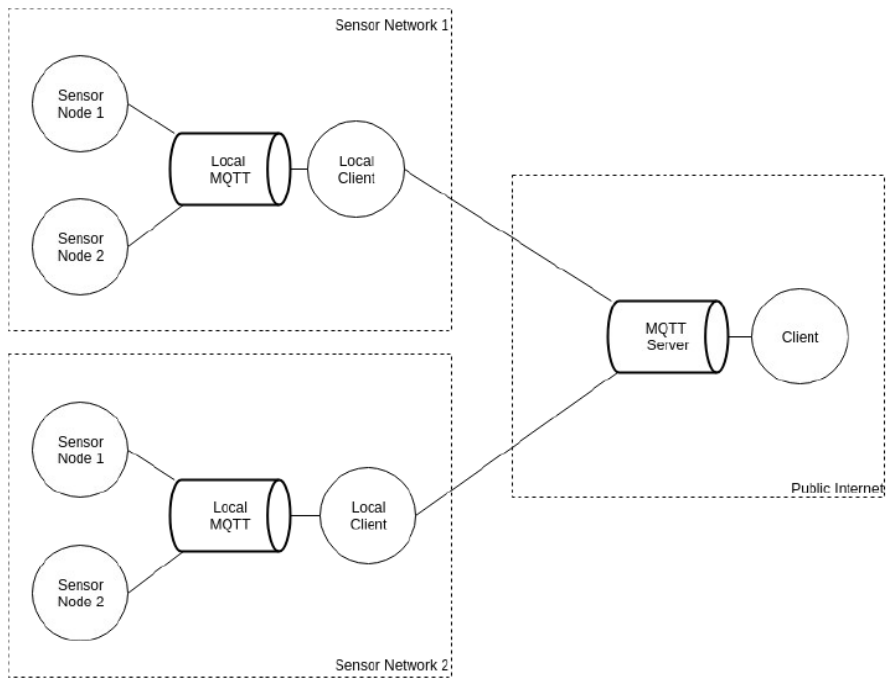

**Fig. 1.** Two-layer MQTT-based network that retransmits data to public cloud

On the other hand, the second of the proposed architectures (Fig. 2) requires sensor nodes to check for connectivity and handle Internet access loss by themselves. This means sensor nodes should keep both connections to local broker and external broker. This also mean sensor nodes should also carry about which data is critical for system operations, and which is not.

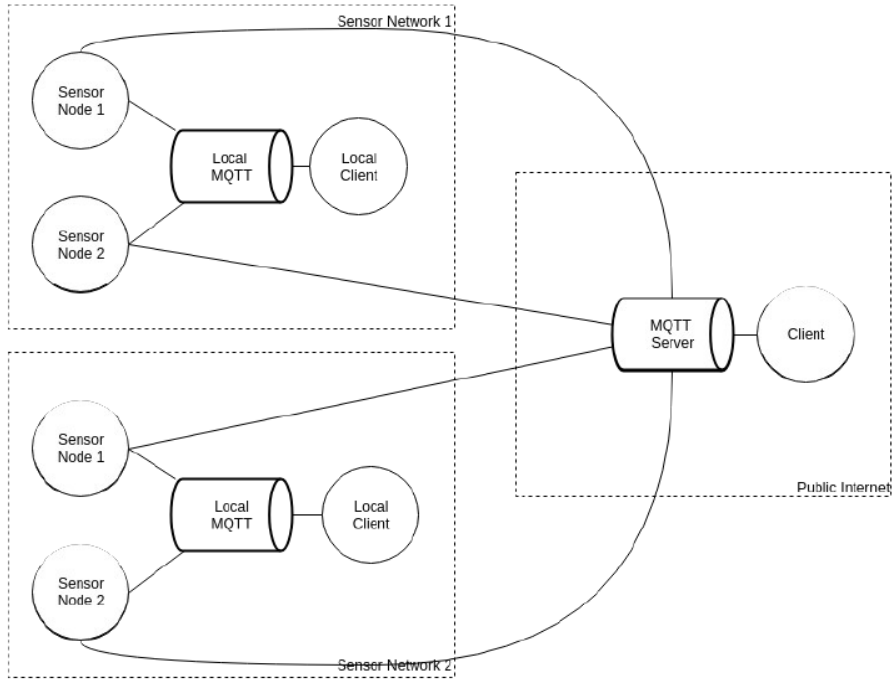

**Fig. 2.** Two-layer MQTT-based network in which sensor nodes handle connectivity checks

### 3.2. Advantages and disadvantages of the proposed architectures

Both of the proposed solutions have the disadvantages of additional software and hardware resources required to support themselves.

As for the advantages of the proposed solutions, they are the following:
1) System autonomous operating support
2) Possibility of faster processing of critical data, as it is processed locally

Besides the above, each of the proposed architectures has its own specifics. These specifics are shown in the Table 1.

**Table 1.** Comparative analysis of proposed architectures

| **Architecture № 1** (Fig. 1) | **Architecture № 2** (Fig. 2) |
|---|---|
| Sensor nodes logic is not changed | Sensor nodes logic is changed |
| Local transmitters are required to keep two connections: both to local and public broker | Low-powered sensors are required to keep two connections: both to local and public broker |
| Due to data retransmission, there are additional delays, when sending data to the public broker | No delays when sending data to public brokers |
| Constant usage of both local and external resources | Possibility offload all the processing only to external resources in case of good Internet connection |
| No duplicated messages in different brokers | There might be duplicated messages in local and public broker |

### 3.3. Message delivery delay investigation

One of the key characteristics of the embedded systems is their response time. Thus, it is necessary to investigate message delivery delays in order to evaluate the effectiveness of the proposed architectures and compare them to the existing ones.

For the research and measurements, we use ESP8266 chip due to its availability and popularity. For this chip the sensor imitation, which generates data and sends them every 5 seconds, is implemented. For the ease of investigation, we use sequential numbers as the payload of messages. This payload is "processed" with the external client, which sends it to the different MQTT topic, to which the sensor is subscribed. Sensor node tracks the time when the message was sent and when the corresponding response has been received. Thus measuring difference between two timestamps will give us the roundtrip time in the system.

In order to eliminate the problem of time synchronization between multiple devices and achieve the maximum accuracy, we perform all measurements with only one chip. Sensor node logic is implemented using C++ programming language and uploaded with platformio tooling.

For time measurements, built-in function *micros()* which returns microseconds since device start is used. As mentioned above, we use QoS 1 (at least once), as this policy gives the optimal balance between message delivery guarantees and protocol data overhead. As QoS 1 allows duplicate messages to be sent/received, we measure only time between sending data and first "response". All the following duplicates are ignored.

For this investigation, we use two MQTT-brokers. External platform io.adafruit.com was used for hosting public MQTT broker. We used eclipse mosquitto (open source implementation of the broker, which supports protocol standards 3.1.1 and 3.1) for setting up local MQTT broker. External platform uses secure connection (over TLS) and user authentication, we do not use those in local broker. This decision is motivated by means of ease of configuration and reducing network overhead as the unauthorized access to the broker can be prevented by properly configuring network itself.

Internet is accessed using LTE connectivity with the download speed of 2.4Mbps and upload speed of 8.1Mbps. These network characteristics were obtained using speedtest.net service. Local connectivity was implemented using Wi-Fi network.

For measurements, we use a simple NodeJS application, which only retransmits messages from one MQTT topic to another, thus giving the minimal impact on delays. In order to retransmit messages from local broker to public and vice-versa another NodeJS application was implemented. The latter accepts two sets of topics: ones that should be retransmitted from local broker to external and ones that should be retransmitted from external broker to local. As MQTT 3.1.1 does not support user properties in messages, there is no way to identify message sender. Due to this fact, sets of topics should not intersect; otherwise a message retransmission flood (infinite retransmission of the same message from local to public and vice-versa) will occur.

## 4    Results

Both architectures was investigated in two different scenarios: in the first scenario there was a constant good connectivity to the Internet and in the second, Internet connection was unavailable (while the local network was still working) for the time of sending a single message. For comparison the same measurements were done for the two single-layer architectures. The first of single layer architecture uses only public MQTT broker (Fig 3a), and thus cannot operate in case of Internet connection loss. The second one uses only a local broker (Fig. 3b) and has no Internet connection at all, thus making it impossible to receive sensor network data from the outside.

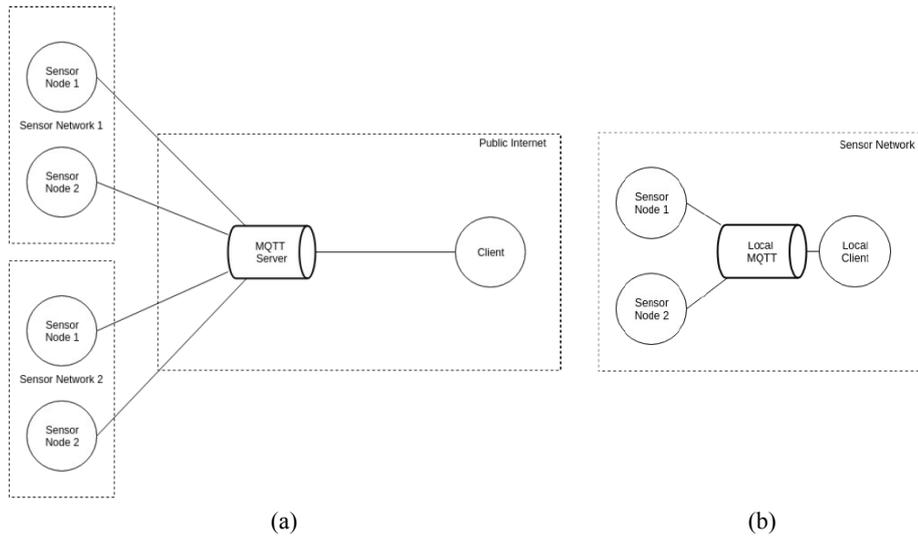

(a)                                        (b)

**Fig. 3**. Single layer architectures with public broker only *(a)* and local broker only *(b)* which are used for comparison.

For each case 20 measurements were performed and average message roundtrip delays were calculated. The results of the measurements (in microseconds) are shown in Table 2.

The comparison of average message delivery delay time is shown on Fig. 4. Visualization allows to compare average delays when delivering messages between the proposed two-layer architectures as well as to compare those to the single-layer architecture.

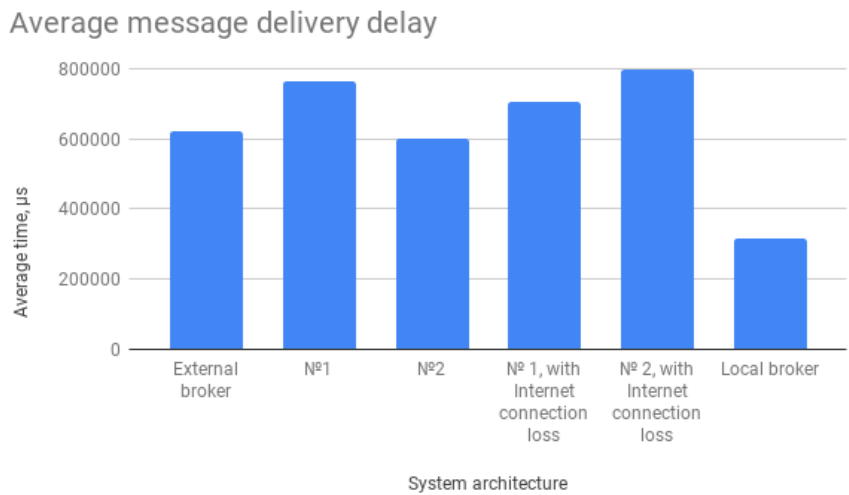

**Fig. 4.** Comparison diagram of average message delivery delay

**Table 2**. Message delivery delays (in microseconds)

| # | System architecture | | | | | |
|---|---|---|---|---|---|---|
| | **External broker** | **№ 1** | **№ 2** | **№ 1, with Internet connection loss** | **№ 2, with Internet connection loss** | **Local broker** |
| 0 | 5.02E+05 | 8.83E+05 | 4.71E+05 | 7.13E+05 | 4.61E+05 | 2.32E+05 |
| 1 | 7.22E+05 | 5.93E+05 | 4.70E+05 | 9.24E+05 | 5.20E+05 | 3.12E+05 |
| 2 | 6.72E+05 | 7.13E+05 | 8.73E+05 | 9.34E+05 | 8.42E+05 | 1.42E+05 |
| 3 | 6.12E+05 | 7.33E+05 | 4.81E+05 | 2.16E+04 | 5.01E+05 | 3.12E+05 |
| 4 | 6.62E+05 | 5.03E+05 | 4.92E+05 | 7.33E+05 | 5.02E+06 | 2.42E+05 |
| 5 | 5.61E+05 | 6.13E+05 | 4.60E+05 | 5.03E+05 | 8.61E+05 | 3.12E+05 |
| 6 | 6.82E+05 | 6.13E+05 | 7.01E+05 | 8.83E+05 | 6.31E+05 | 2.12E+05 |
| 7 | 5.61E+05 | 9.34E+05 | 5.02E+05 | 8.14E+05 | 4.70E+05 | 3.02E+05 |
| 8 | 6.01E+05 | 8.14E+05 | 5.61E+05 | 7.33E+05 | 5.61E+05 | 1.92E+05 |
| 9 | 9.63E+05 | 8.34E+05 | 4.91E+05 | 7.64E+05 | 4.61E+05 | 5.13E+05 |
| 10 | 5.91E+05 | 7.84E+05 | 5.01E+05 | 5.23E+05 | 4.71E+05 | 5.13E+05 |
| 11 | 5.01E+05 | 6.84E+05 | 6.11E+05 | 8.34E+05 | 4.41E+05 | 5.13E+05 |
| 12 | 6.31E+05 | 5.93E+05 | 5.51E+05 | 5.93E+05 | 5.11E+05 | 4.12E+05 |
| 13 | 5.31E+05 | 5.23E+05 | 5.83E+05 | 7.84E+05 | 5.90E+05 | 1.42E+05 |
| 14 | 7.22E+05 | 9.34E+05 | 5.51E+05 | 6.13E+05 | 4.84E+05 | 4.16E+04 |
| 15 | 5.52E+05 | 9.24E+05 | 4.91E+05 | 9.24E+05 | 4.91E+05 | 3.84E+05 |
| 16 | 6.92E+05 | 7.64E+05 | 7.83E+05 | 6.13E+05 | 1.01E+06 | 3.02E+05 |
| 17 | 5.72E+05 | 9.24E+05 | 7.81E+05 | 5.93E+05 | 5.02E+05 | 5.13E+05 |
| 18 | 5.01E+05 | 7.33E+05 | 8.81E+05 | 6.84E+05 | 6.32E+05 | 2.32E+05 |
| 19 | 6.31E+05 | 1.18E+06 | 7.92E+05 | 9.34E+05 | 4.70E+05 | 5.23E+05 |
| **Total** | 1.25E+07 | 1.53E+07 | 1.20E+07 | 1.41E+07 | 1.59E+07 | 6.35E+06 |
| **Avg** | 6.23E+05 | 7.64E+05 | 6.01E+05 | 7.06E+05 | 7.97E+05 | 3.17E+05 |
| **Stdev** | 1.07E+05 | 1.70E+05 | 1.44E+05 | 2.12E+05 | 1.01E+06 | 1.44E+05 |

Having standard deviation calculation we can investigate actual measurements deviation from the average value (Fig. 5).

Average message delivery delay time comparison to the ones of one-layered architecture are shown in Table 3

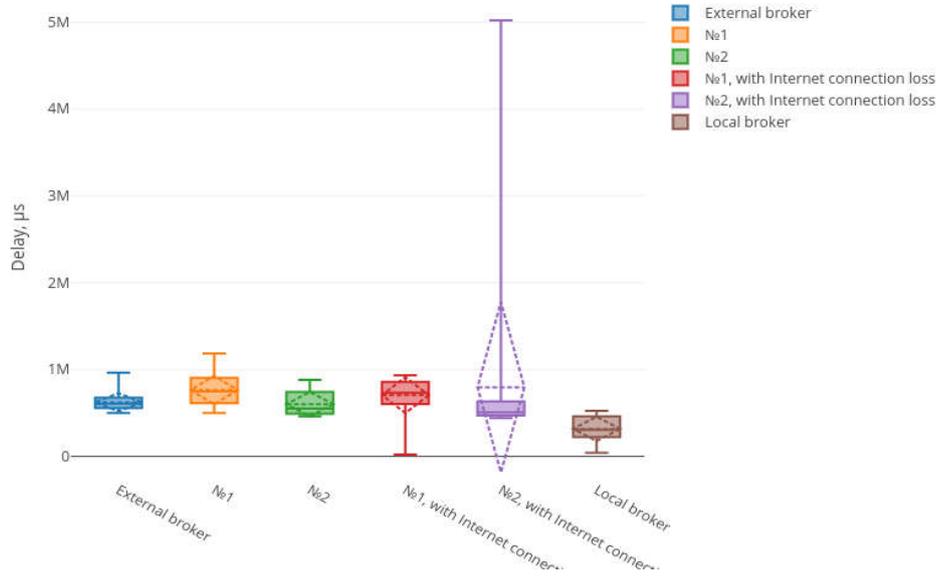

**Fig. 5.** Statistical analysis of the observed results

**Table 3**. Message delivery delays comparison

|  | External broker | #1 | #2 | #1, Internet connection loss | #2, Internet connection loss | Local broker |
|---|---|---|---|---|---|---|
| **Average time,** | 6.23E+05 | 7.64E+05 | 6.01E+05 | 7.06E+05 | 7.97E+05 | 3.17E+05 |
| **Compared to external broker,** | 0 | 1.41E+05 | -2.18E+04 | 8.28E+04 | 1.74E+05 | -3.06E+05 |
| **Compared to a local broker,** | 3.06E+05 | 4.47E+05 | 2.84E+05 | 3.89E+05 | 4.80E+05 | 0 |

The proposed architectures solve the problem of both online and offline work of the sensor network by adding an additional layer, which consists of local MQTT broker. Proposed solution #1 does not require sensor logic modification and solution #2 allows offloading computations to external network completely if it is available.

## 5 Discussion

Results demonstrate (see Table 2) that message delivery delays are the lowest for the architecture with the local broker only. This approach cannot always be used, as it does not provide the possibility to access data from outside the sensor network.

As for the delays when Internet connection loss occurs, average delay time is increased in both two-layer architectures. This is caused by the implementation details: by default, connection status is checked every 5 seconds and this time was spent until connection loss was noticed and system switched to the local broker. Connectivity check timeout can be configured, thus opening the field for the further investigations.

The average message delay time decrease, which was observed using proposed architecture #2, was insignificant and probably caused by the mobile network load.

The research shown that proposed architecture #2 does not have overhead when stable Internet connection is available compared to the external broker architecture. However, a significant delay occurs when Internet connection is lost.

Based on the observations the effective solution for practical application can be proposed. The smallest delays will be achieved by moving most of computations and interactions to the local network with communicating using local MQTT broker. All data, which is required outside the sensor network can be retransmitted to external broker, as well as all commands from the external network can be retransmitted to the local MQTT as in the proposed architecture № 1. In general, these results support the general trend of IoT and CoT integration in the wider context of Cloud-Fog-Dew computing paradigm [4-5].

## 6 Conclusions

The purpose of this work was to develop a solution for system components in the context of the Internet of Things that provide the ability to operate such systems in both online and offline modes. The alternative two layer architectures were proposed and investigated. Based on the investigation results, the more optimal two-layer architecture was proposed.